\documentclass[12pt]{article}
\usepackage{amsmath}
\usepackage{amssymb}
\usepackage{amsthm}
\usepackage{amsfonts}
\usepackage{graphicx}
\usepackage{textcomp}
\usepackage{color}
\usepackage{hyperref}
\usepackage{tikz-cd}
\numberwithin{equation}{section}

\begin{document}

\begin{titlepage}

\title{Weyl Covariant Theories of Gravity in Riemann-Cartan-Weyl Space-times \\ II. Minimal Massive Gravity}

\author{ Tekin Dereli\footnote{tdereli@ku.edu.tr}, Cem Yeti\c{s}mi\c{s}o\u{g}lu\footnote{cyetismisoglu@ku.edu.tr} \\ {\small Department of Physics, Ko\c{c} University, 34450 Sar{\i}yer, \.{I}stanbul, Turkey }}

\date{25 April 2019}

\maketitle

\begin{abstract}
\noindent We present locally scale (Weyl) covariant generalisation of Minimal Massive Gravity theory using the language of exterior differential forms on Riemann-Cartan-Weyl space-times. The theory is expressed by a locally scale invariant action and locally scale covariant field equations are found by a first order variational formalism.
\end{abstract}

\vskip 2cm

\noindent {\bf Keywords}: Riemann-Cartan-Weyl Spaces $\cdot$ Scale Invariance $\cdot$ Massive Gravity Theories

\end{titlepage}

\maketitle			
\clearpage 

\section{Introduction}
In an earlier study, we discussed a method for achieving locally scale covariant generalisations of gravitational theories using Riemann-Cartan-Weyl (RCW) space-times [15]. RCW space-time is a geometry that is equipped with the most general linear connection. That is, in addition to the metric, the geometry has both torsion and non-metricity. The method of gauging the Weyl group involves introducing a real scalar field and a connection 1-form for the Weyl group. The connection 1-form of Weyl group is identified with the trace part of non-metricity tensor. This identification geometrizes the origin of scale transformations. In [15], we showed this procedure provides a consistent generalisation for General Relativity (GR) and Topologically Massive Gravity (TMG) theories. In this paper, we focus on the Minimally Massive Gravity (MMG) theory in three dimensions (3D). \\

\noindent 3D gravity theories are important because they provide important toy models in pursuit of the quantum gravity problem [1, 2, 3]. It is well known that GR theory in 3D is locally trivial, i.e. there is no propagating degrees of freedom [4]. As the Einstein tensor is proportional to the dualised Riemann tensor, it admits only spaces of constant curvatures as solutions. However, by virtue of 3D, one can couple gravitational Chern-Simons term to the action and obtain a dynamical theory called TMG theory [5, 6]. Although having high derivative degrees, when linearized around an $AdS_3$ background, the theory have unitary massive spin-2 field excitations for Einstein-Hilbert term having the wrong sign. This problem is called the bulk versus boundary clash problem. An alternative theory that modifies 3D GR and solves this problem is MMG theory [7, 8]. As an extension of TMG, this model has the same degrees of freedom as TMG. But for positive energy bulk modes, this theory has positive central charge for the dual CFT at the boundary of $AdS_3$ background. In addition, MMG theory has dynamical space-time torsion present in it, and is different from TMG in this respect. \\

\noindent Although it is a natural generalisation, scale covariant MMG theory has not been studied before. Due to dynamical space-time torsion present in MMG field equations, it is harder to generalise it to a scale covariant theory.  Scale covariant generalisations are useful for discussing high and low energy complete theories. Also, such a generalisation may be supplemented with a mass generation scheme for the massive graviton mode via a Higgs like mechanism [9]. Furthermore, one can look for solutions of the scale covariant theory that are Einstein-Weyl spaces. 3D Einstein-Weyl spaces are very rich [10, 11] and generalise 3D Einstein spaces. 3D Einstein-Weyl spaces are classified by Cartan [12] and are parametrised by eight real parameters whereas 3D Einstein spaces are parametrised by only a single parameter (that is the cosmological constant). In addition, 3D Einstein-Weyl spaces have equivalent formulations in terms of twistor spaces via the Hitchin correspondence [13]. Furthermore, these twistor spaces can be used to construct four dimensional self-dual geometries using the Penrose correspondence [14].  \\

\noindent In our work, we use the language of exterior algebra on RCW space-times. We follow exactly the conventions in [15]. For the formulation of MMG theory using exterior algebra on Riemann-Cartan geometries, we refer to [8]. Now we formulate the scale covariant MMG theory using an action principle. The field equations are found using a first order variational formalism. We then show the consistency of the generalisation. Finally, we finish with concluding remarks. 

\section{Locally Weyl Covariant Generalisation of Minimal Massive Gravity Theory}

\noindent The locally scale covariant MMG theory is going to be defined using an action principle. The field equations are going to be derived using a first order variational formalism. We turn MMG into a locally scale covariant theory by introducing a real scalar field $\alpha$ of scale charge 1. The real field $\alpha$ is called the dilaton field and can be viewed as inverse of the local gravitational coupling strength. Furthermore, we take into most general linear connection 1-form with its symmetric part $Q^a_{\ b}$ identified with the Weyl connection 1-form $Q$ as:
\begin{equation}
\Lambda^a_{\ b}=\Omega^a_{\ b}+Q^a_{\ b}, \quad Q^a_{\ b}= -\eta^a_{\ b}Q
\end{equation}
where $\Omega^a_{\ b}$ denote the anti-symmetric part of the connection. Consequently, variations of symmetric part of connection 1-forms $\{Q^a_{\ b}\}$ are related to the variations of Weyl connection 1-form $Q$. The independent variables with respect to which the Lagrangian 3-form will be varied are: co-frame 1-forms $\{e^a\}$, anti-symmetric part of connection 1-forms $\{\Omega^a_{\ b}\}$, Weyl connection 1-form $Q$, dilaton field $\alpha$ and Lagrange multiplier 1-forms $\{\lambda_a\}$. \\

\noindent For the locally Weyl covariant MMG theory, the action functional will be given by 
\begin{equation}
I[e^a, \Omega^a_{\ b}, Q, \alpha, \lambda_a ] = \int_M \mathcal{L}
\end{equation}
where $M$ is a compact region without boundary on a (2+1)-dimensional RCW manifold. The Weyl invariant Lagrangian density for the theory is given by:
\begin{align}
\mathcal{L}&=\frac{1}{\mu}(\Lambda^a_{\ b} \wedge d\Lambda^b_{\ a} + \frac{2}{3} \Lambda^a_{\ b} \wedge \Lambda^b_{\ c} \wedge \Lambda^c_{\ a})+\frac{1}{\mu'}Q \wedge dQ +\alpha \overset{(\Lambda)}{R^a_{\ b}} \wedge *e_a^{\ b}+ \alpha^3\Lambda*1    \nonumber\\
&-\frac{\gamma}{2\alpha} \mathcal{D} \alpha \wedge * \mathcal{D} \alpha-\frac{\gamma^{\prime}}{2\alpha}dQ \wedge *dQ +\alpha\lambda_a\wedge T^a + \frac{\nu\alpha}{2} \lambda_a \wedge \lambda_b \wedge *e^{ab}. 
\end{align}
Above, the coupling constants $\mu$, $\mu'$, $\gamma$, $\gamma'$ and $\nu$ are dimensionless due to Weyl invariance. Also, we added the kinetic term of the dilaton field to promote it to a dynamical field. \\

\noindent Weyl covariant field equations are found by taking the variational derivative of the Lagrangian density 3-form which reads (up to a closed form):
\begin{align}
\dot{\mathcal{L}} &= {\dot{e}}^a \wedge \bigg\{ \alpha \overset{(\Omega)}{R^b_{\ c}} \epsilon_{ab}^{\ \ c} +\alpha^3\Lambda *e_a + \overset{(\Omega)}{D}(\alpha\lambda_a)+\alpha Q \wedge \lambda_a +\frac{\gamma}{2\alpha}\tau_a[\mathcal{D}\alpha]+\frac{\gamma'}{2\alpha}\hat{\tau}_a[dQ]\nonumber\\
&+\frac{\nu\alpha}{2}\epsilon_a^{\ bc}\lambda_b \wedge \lambda_c  \bigg\} + {\dot{\Omega}}^a_{\ b} \wedge \bigg\{ \frac{2}{\mu}\overset{(\Omega)}{R^b_{\ a}} + \overset{(\Omega)}{D}( \alpha *e_a^{\ b}) +\alpha e^b \wedge \lambda_a \bigg\}\nonumber\\
& +\dot{\lambda_a} \wedge \bigg\{\alpha (T^a+\nu\lambda_b \wedge *e^{ab})\bigg\}+\dot{\alpha}\bigg\{\overset{(\Omega)}{R^a_{\ b}} \wedge *e_a^{\ b} +3\alpha^2 \Lambda*1 +\frac{\gamma}{2\alpha^2}\mathcal{D}\alpha \wedge*\mathcal{D}\alpha \nonumber\\
&+\gamma\mathcal{D}\bigg(\frac{1}{\alpha}*\mathcal{D}\alpha\bigg)+\frac{\gamma'}{\alpha^2}dQ \wedge *dQ +\lambda_a \wedge T^a +\frac{\nu}{2} \lambda_a \wedge \lambda_b \wedge *e^{ab}\bigg\} \nonumber\\
& +\dot{Q} \wedge \bigg\{\bigg(\frac{6}{\mu}+\frac{2}{\mu'}\bigg)dQ+ \alpha \lambda_a \wedge e^a- \gamma*\mathcal{D}\alpha-\gamma' d\bigg(\frac{1}{\alpha}*dQ\bigg) \bigg\}  \label{lag}
\end{align}
where a dot over a field variable means its variation and the expressions
\begin{align}
\tau_a[\mathcal{D}\alpha]&=(\iota_a\mathcal{D}\alpha)*\mathcal{D}\alpha+\mathcal{D}\alpha \wedge \iota_a *\mathcal{D}\alpha, \\
\hat{\tau}_a[dQ]&=\iota_adQ \wedge *dQ-dQ (\iota_a *dQ),
\end{align}
are the scale covariant stress-energy forms of the dilaton field $\alpha$ and the Weyl vector boson field $Q$, respectively. \\

\noindent Now, we begin to simplify the variational field equations. We start by solving the Lagrange multiplier equation for torsion 2-forms. 
\begin{equation}
T^a+\nu\lambda_b \wedge *e^{ab}=0 \Leftrightarrow T^a = - \nu\lambda_b \wedge *e^{ab}. \label{tor}
\end{equation}
Therefore, the auxiliary field 1-forms are proportional to dualised contorsion 1-forms and the rest of the field equations will be solved under this constraint. The dilaton field equation can be simplified greatly. For this, we first take the trace of co-frame equation. Trace of the co-frame equation follows from left exterior multiplication with $e^a$ which reads:
\begin{align}
&\alpha\overset{(\Omega)}{R^a_{\ b}} \wedge *e_a^{\ b} +3\alpha^2 \Lambda*1 -\frac{\gamma}{2\alpha}\mathcal{D}\alpha \wedge*\mathcal{D}\alpha+\frac{\gamma'}{2\alpha}dQ \wedge *dQ  \nonumber\\
&+\alpha \lambda_a \wedge T^a+\frac{\nu\alpha}{2} \lambda_a \wedge \lambda_b \wedge *e^{ab}  - d (\alpha e_a \wedge \lambda^a )=0.
\end{align}
Comparing this equation with the dilaton field equation, we obtain a much simpler equation for the dilaton field:
\begin{equation}
d(\alpha e_a \wedge \lambda^a+\gamma*\mathcal{D}\alpha)=0.
\end{equation} 
Next, using
\begin{equation}
\overset{(\Omega)}{D}(\alpha*e_a^{\ b})=\mathcal{D}\alpha \wedge *e_a^{\ b} + \alpha \epsilon_{ca}^{\ \ b}T^c,
\end{equation}
together with the expression for torsion 2-forms (\ref{tor}), we rewrite the anti-symmetric part of the connection equation as:
\begin{equation}
M(e_a\wedge \lambda_b - e_b \wedge \lambda_a)=\Sigma_{ab}, \label{ancon}
\end{equation}
where the shorthand expressions read:
\begin{equation}
M= \frac{\alpha}{2} -\alpha\nu, \quad \Sigma_{ab}=-\frac{2}{\mu}\overset{(\Omega)}{R_{ab}}+\mathcal{D}\alpha \wedge *e_{ab}.
\end{equation}
From equation (\ref{ancon}), auxiliary field 1-forms $\{\lambda_a\}$ can be solved uniquely as:
\begin{equation}
\lambda_a = -\frac{4}{\mu\alpha(1-2\nu)}\overset{(\Omega)}{Y_a}+ \frac{2}{\alpha(1-2\nu)} \iota_a*\mathcal{D}\alpha.
\end{equation}
Above, the Schouten 1-forms $\{\overset{(\Omega)}{Y_a}\}$ of the anti-symmetric part of connection 1-forms $\{\Omega^a_{\ b}=\omega^a_{\ b}+K^a_{\ b}+q^a_{\ b}\}$ are defined as:
\begin{equation}
\overset{(\Omega)}{Y_a}= \iota^b\overset{(\Omega)}{R_{ba}}-\frac{1}{4}(\iota^{bc}\overset{(\Omega)}{R_{cb}})e_a.
\end{equation}
Finally, the variational field equations of the scale covariant MMG theory read:
\begin{align}
&-2\alpha \overset{(\Omega)}{G_a} +\alpha^3\Lambda *e_a +\frac{\gamma}{2\alpha}\tau_a[\mathcal{D}\alpha] +\frac{\gamma'}{2\alpha}\hat{\tau}_a[dQ]  -\frac{4}{\mu(1-2\nu)}\bigg(\overset{(\Omega)}{D}\overset{(\Omega)}{Y_a}+Q\wedge\overset{(\Omega)}{Y_a}\bigg) \nonumber\\
&+\frac{2}{1-2\nu}\bigg(\overset{(\Omega)}{D}(\iota_a*\mathcal{D}\alpha)+Q \wedge (\iota_a*\mathcal{D}\alpha) \bigg) +\frac{2\nu }{\alpha(1-2\nu)}\epsilon_a^{\ bc}\bigg[\frac{4}{\mu^2} \overset{(\Omega)}{Y_b} \wedge \overset{(\Omega)}{Y_c} \nonumber\\
& -\frac{2}{\mu} \bigg( \overset{(\Omega)}{Y_b} \wedge \iota_c *\mathcal{D}\alpha - \overset{(\Omega)}{Y_c} \wedge \iota_b *\mathcal{D}\alpha \bigg)+ \iota_b *\mathcal{D}\alpha \wedge \iota_c *\mathcal{D}\alpha \bigg]  = 0, \label{wmmg1}
\end{align}
\begin{equation}
\bigg(\gamma + \frac{4}{1-2\nu}\bigg) *\mathcal{D}\alpha= \bigg(\frac{6}{\mu}+\frac{2}{\mu'} - \frac{4}{\mu(1-2\nu)} \bigg) dQ-\gamma'd\bigg(\frac{1}{\alpha}*dQ\bigg),  \label{wmmg2}
\end{equation}
\begin{equation}
\bigg(\gamma + \frac{4}{1-2\nu}\bigg) d*\mathcal{D}\alpha=0. \label{wmmg3}
\end{equation}
To show that this generalisation is consistent, we show that original field equations of MMG theory lie in the vacuum configuration of Weyl sector. To this end, we make the choice 
\begin{equation}
\mathcal{D}\alpha=0 \ \     \Leftrightarrow  \ \ Q=-\frac{d\alpha}{\alpha}.
\end{equation}
Therefore the Weyl connection 1-form is a pure gauge and the Weyl field strength 2-form vanishes, that is $dQ=0$. This is what we mean by vacuum class of solutions for the Weyl sector of the locally scale covariant MMG theory. Also, connection 1-forms are still not metric compatible. For the vacuum configuration, the field equations (\ref{wmmg2}) and (\ref{wmmg3}) vanish identically; and the Einstein field equations (\ref{wmmg1}) reduce to:
\begin{equation}
-2\alpha \overset{(\Omega)}{G_a} +\alpha^3\Lambda *e_a  -\frac{4}{\mu(1-2\nu)} \bigg(\overset{(\Omega)}{D}\overset{(\Omega)}{Y_a}-\frac{d\alpha}{\alpha} \wedge\overset{(\Omega)}{Y_a} \bigg)+\frac{8\nu}{\alpha\mu^2(1-2\nu)}\epsilon_a^{\ bc} \overset{(\Omega)}{Y_b} \wedge \overset{(\Omega)}{Y_c} =0.
\end{equation}
At this point, we still have a residual gauge freedom for the dilaton field $\alpha$. Fixing this residual gauge freedom by setting $\alpha=1$ amounts to choosing a constant fiducial units system to the theory. Besides, after making this choice, Weyl connection 1-form vanishes, i.e. $Q=0$ and connection 1-forms become the metric compatible connection 1-forms $\{\Omega^a_{\ b}=\omega^a_{\ b}+K^a_{\ b} \}$. Then, the Einstein field equations reduce to the original MMG field equations:
\begin{equation}
-2 \overset{(\Omega)}{G_a} +\Lambda *e_a  -\frac{4}{\mu(1-2\nu)}\overset{(\Omega)}{C_a}+\frac{8\nu}{\mu^2(1-2\nu)}\epsilon_a^{\ bc} \overset{(\Omega)}{Y_b} \wedge \overset{(\Omega)}{Y_c} =0
\end{equation}
where the Cotton 2-form of our metric compatible connection is defined in terms of Schouten 1-form as:
\begin{equation}
\overset{(\Omega)}{C_a}:= \overset{(\Omega)}{D}\overset{(\Omega)}{Y_a}.
\end{equation}
This calculation demonstrates that Weyl covariant MMG theory contains the original theory in the vacuum configuration of its Weyl sector. Hence, scale covariant MMG theory defined by the Lagrangian (\ref{lag}) is a consistent generalisation.

\section{Concluding Remarks}
In this study, we considered the locally scale covariant generalisation of MMG theory. Geometry of the theory is a RCW space-time with a non-metricity tensor that has only trace part. The trace part of the non-metricity tensor is identified with the Weyl connection 1-form so that scale transformations become a piece of the space-time geometry. We formulated the theory using an action principle. The scale covariant field equations are found using a first order variational formalism. Afterwards we showed the consistency of this generalisation. Due to dynamical space-time torsion present in MMG theory, it is a challenging task to obtain a scale covariant generalisation. However, this task becomes rather clear by using the compelling language of exterior algebra. \\

\noindent Due to scale covariance, there are no dimensionful parameters in the theory. Therefore, mass of the gravitons occur after the scale symmetry is broken. For further studies, one can supplement the dilaton field with a potential and generate a Higgs-like symmetry breaking. This way one can gain insight on how the gravitons obtain their masses. Besides, one can look for non-trivial solutions of the scale covariant theory that are Einstein-Weyl spaces. Thereon one can incorporate Hitchin and Penrose correspondences and can generate new space-time models. Alternatively, one can directly solve the field equations, and look for a stable vacuum and investigate the particle spectrum. We believe our results will be helpful for investigating these questions.

\end{document}